# Grid-independent Issue in Numerical Heat Transfer

Yao Wei，Wang Jian*, Liao Guangxuan

(State Key Lab. of Fire Science, University of Science and Technology of China, Hefei, 230027, China)

**Abstract**: Grid independent is associated with the accuracy or even rationality of numerical results. This paper takes two-dimensional steady heat transfer for example to reveal the effect of grid resolution on numerical results. The law of grid dependence is obtained and a simple mathematical formula is presented. The production acquired here can be used as the guidance in choosing grid density in numerical simulation and get exact grid independent value without using infinite fine grid. Through analyzing grid independent, we can find the minimum number of grid cells that is needed to get grid-independent results. Such strategy can save computational resource while ensure a rational computational result.

## 0 Introduction

Grid independent is a very important issue in numerical simulation, but still now little literatures has concerned with this matter. Grid-independent means calculational results change so little along with a denser or looser grid that the truncation error can be ignored in numerical simulation. Whether the grid is independent directly influences the truncation error or even the rationality of numerical results. The previous research reveals that the grid resolution and time step have a very great effect on the results of unsteady numerical simulation in a certain range[1]. When considering grid-independent issue, in principle a very dense grid can avoid this problem but the calculational resource may be wasted unnecessarily. In practice, we usually increase the grid resolution according to a certain ratio, for example 1/3, and then compare the results of two neighborhood results[2]. If the results tend towards identical, the grid can be considered as grid-independent. Such strategy can utilize computational resource most efficiently as well as obtain reasonable results.

The truncation error caused by grid resolution is defined as the difference between the grid independent result $\phi_{grid\ indep.}$ and current numerical result $\phi$:

$$\varepsilon = \phi_{grid\ indep.} - \phi \tag{1}$$

According to previous studies, truncation error $\varepsilon$ will become smaller and smaller by refined computational grid[3, 4]. But in what way $\phi \to \phi_{grid\ indep.}$ is still unclear. This paper takes two-dimensional steady heat transfer for example to illustrate how numerical result $\phi$ converges towards $\phi_{grid\ indep.}$ when refining computational grid.

# 1 Mathematical Model

Here we only consider conduction when referring to heat transfer. Heat conduction is the transfer of heat from warm areas to cooler ones, and effectively occurs by diffusion[5]. Heat conduction follows the Fourier's law

$$\Phi_Q = -k\nabla T\,, \tag{2}$$

as a result of which temperature changes in time and space. Where $\Phi_Q$ is heat flux, $k$ is thermal conductivity.

For two-dimensional steady heat transfer, assuming thermal diffusivity is constant, the heat transfer equation is:

$$\frac{\partial^2 t}{\partial x^2} + \frac{\partial^2 t}{\partial y^2} = S \tag{3}$$

with appropriate boundary conditions. S stands for heat source and if there is no heat source, S=0.. This equation is Laplace equation or Poisson equation in mathematical angle of view[6]. We can obtain the temperature in the entire field when the value of the function is specified on all boundaries (Dirichlet boundary conditions, fixed temperature) or the normal derivative of the function is specified on all boundaries (Neumann boundary conditions, adiabatic or specified heat flux).

# 2 Numerical Methods

Since the steady heat transfer equation is a conservative full-potential equation, we can adopt AFI methods to solve it. The advantage of the AFI method is it can converge quickly and more accurate[7].

For stable heat transfer equation(), we have the general iterative scheme[8]

$$N\Delta_{i,j}^n = L\phi_{i,j}^n \tag{4}$$

Where L is a finite difference operator.

The usual second accurate central difference approximation is

$$L\phi_{i,j}^n = \left( \frac{\overrightarrow{\delta}_x \overleftarrow{\delta}_x}{\Delta x^2} + \frac{\overrightarrow{\delta}_y \overleftarrow{\delta}_y}{\Delta y^2} \right) \phi_{i,j}^n - S \tag{5}$$

The second order central difference operator $\overrightarrow{\delta}_x \overleftarrow{\delta}_x$ has been formed by a combination of the first order forward and backward difference operator in the $x$ direction. These are defined by

$$\overrightarrow{\delta}_x \phi_{ij} = \phi_{i+1,j} - \phi_{ij} \quad (6)$$

$$\overleftarrow{\delta}_x \phi_{ij} = \phi_{i,j} - \phi_{i-1,j} \quad (7)$$

We define the correction vector for cycle $n+1$ as

$$\Delta_{i,j}^{n} = \phi_{i,j}^{n+1} - \phi_{i,j}^{n} . \quad (8)$$

So the AFI scheme for heat transfer full-potential equation is

$$(\alpha - \frac{\overrightarrow{\delta}_x \overleftarrow{\delta}_x}{\Delta x^2})(\alpha - \frac{\overrightarrow{\delta}_y \overleftarrow{\delta}_y}{\Delta y^2})\Delta_{i,j}^{n} = 2\alpha L\phi_{i,j}^{n} \quad (9)$$

Acceleration parameter $\alpha$ can be chosen according to the parameter sequence proposed by W.F.BALLHAUS et al[9], which has proved to be very effective.

Orthogonal grid based on Cartesian coordinate is adopted here.

It should be pointed that most of the studies in numerical simulation did not focus on minimizing the effect of the numerical errors. If care is not taken to reduce the numerical errors, they may be larger than the contribution from the grid resolution itself. Therefore, it will not be possible to separate the numerical effects from the performance of grid. The use of AFI method clearly reduces the numerical errors in the simulation when compared to the studies previously mentioned. According to the study by Zhu[7], the AFI method used ensures that the contribution from the grid resolution is larger than the numerical errors. Thus, the use of AFI method creates a numerically-clean environment where the effect of grid on numerical results can be tested and validated.

# 3 Results and Discussion

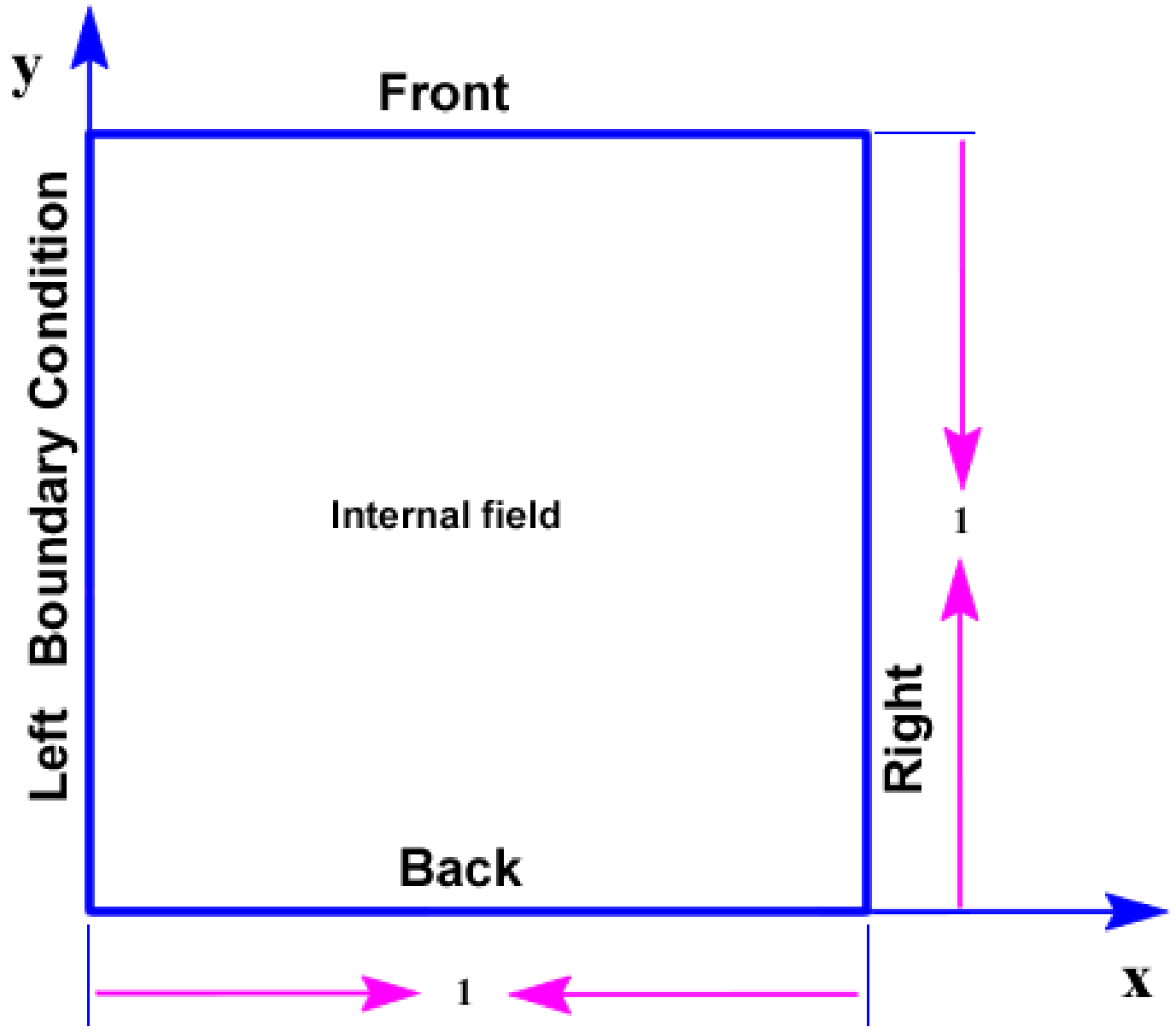

Fig. 1 Schematic diagram of heat transfer with appropriate boundary conditions

Fig.1 shows the physical model of numerical simulation. Results are analyzed when grid resolution varies from coarse to fine. Below when we refer to number of grid cells $n$, we mean the grid on the whole field is $n \times n$.

## 3.1 Simulation of heat transfer with heat source

At all boundaries temperature is set to fixed value 0. The whole field has a uniform heat source. Fig.2 shows that temperature distribution of field with heat source. The temperature is highest in the center, decreasing gradually towards boundaries, which displays a convex shape in the temperature field.

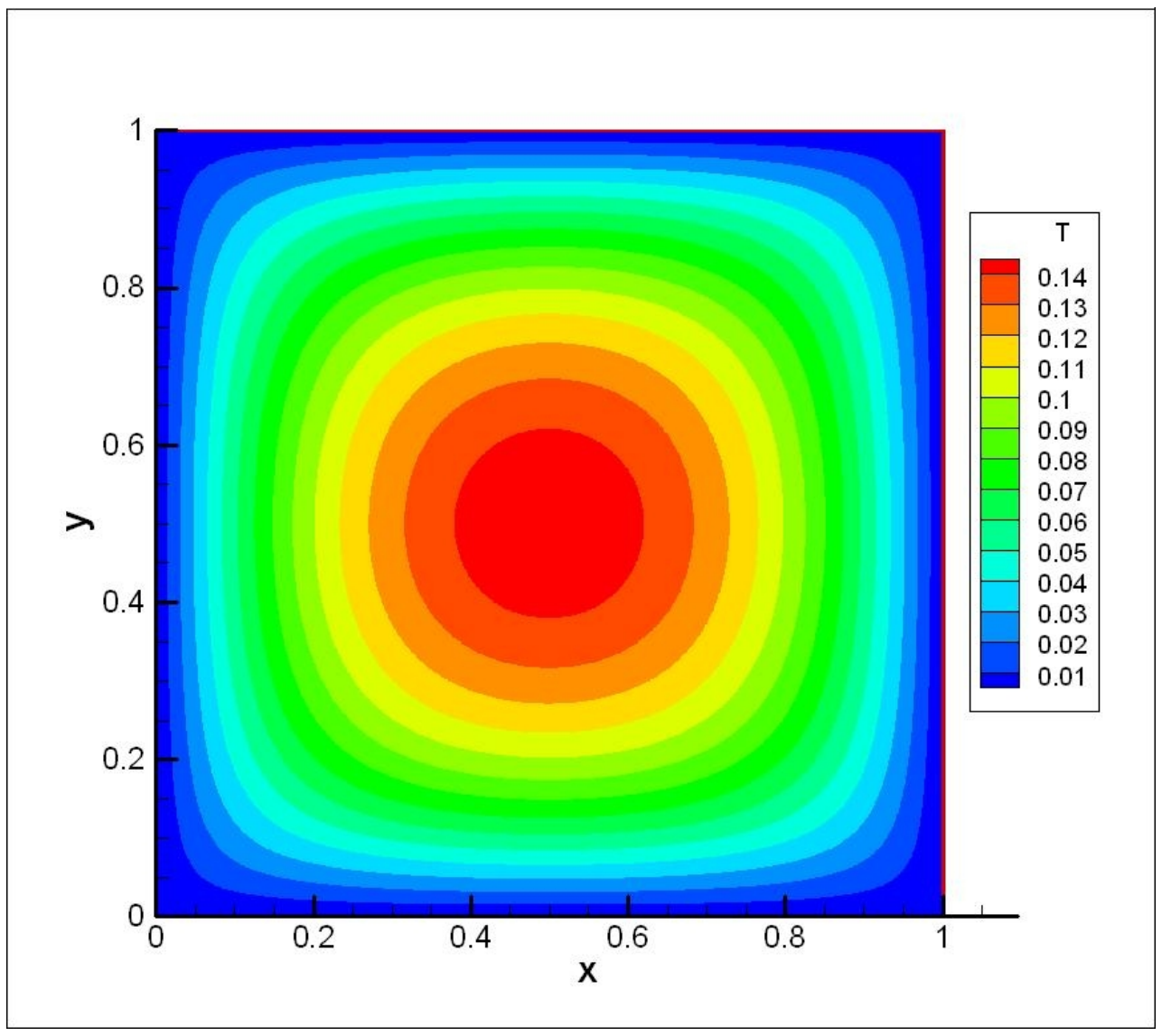

Fig. 2 Temperature field of heat transfer with heat source; Boundaries are set to fixed temperature

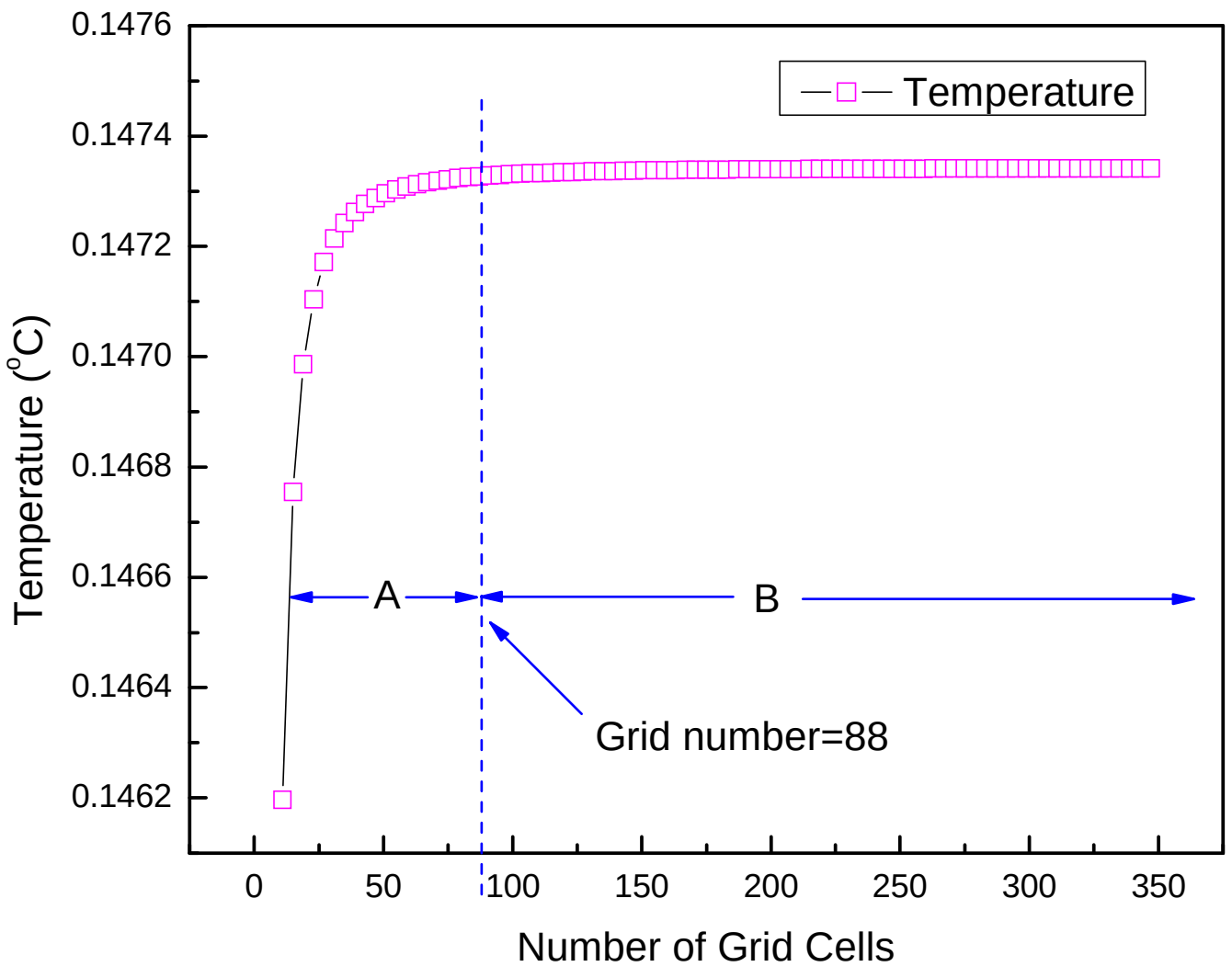

Fig.3 Curve of center temperature against number of grid cells

Fig. 3 is the curve of center temperature against grid resolution, which shows the effect of grid in numerical heat transfer. Actually at each net node the curve of temperature against grid resolution resembles the curve in Fig. 3. From Fig.3 we can see that the variation of temperature against grid resolution obeys its particular law. As in Fig.3, at stage A the results vary with grid resolution greatly; at stage B, the results tend towards constant, so the grid at stage B is grid-independent. In this study, the divided grid number is 88, any grid number more than 88 can be considered as grid-independent.

It is found that the curve like in Fig, 3 accords with such a mathematical formula:

$$y = f(x) = a + \frac{b}{x} + \frac{c}{x^2} + \cdots + \frac{q}{x^n} = \sum_{i=0}^{n} a_i \frac{1}{x^i} \qquad (10)$$

A common sense in numerical simulation is that when increasing grid resolution, the numerical results will tend toward identical and become a constant. This has been validated by previous numerical studies. In true LES, the solution is obtained by keeping the filter width constant while the computational grid is refined[10]. As the grid is refined, the solution converges towards the true solution. The true solution will depend on the mathematical methods used, but will be independent of the grid resolution. This can be obviously seen from the form of formula (10).

$$\lim_{x\to\infty}\left(\frac{a}{x}\right)^i = 0 \qquad (i = 1\cdots n) \qquad (11)$$

So

$$\lim_{x\to\infty} f(x) = a \qquad (12)$$

$a$ is the grid independent value in numerical simulation. Through (10), The grid independent numerical results are obtained.

Generally, taking $n = 3$ is enough to model the curve of numerical results with respect to grid resolution. The curve in Fig. 3 can be expressed as such a mathematical form:

$$\text{y=f(x)=0.147343-}\frac{0.260054}{\text{x}^3} - \frac{0.114794}{x^2} - \frac{0.0000194606}{x} \qquad (11)$$

The grid independent value of temperature at central point is 0.147343.

As shown in Fig. 4, we can see that the agreement between the mathematical formula and data is perfect.

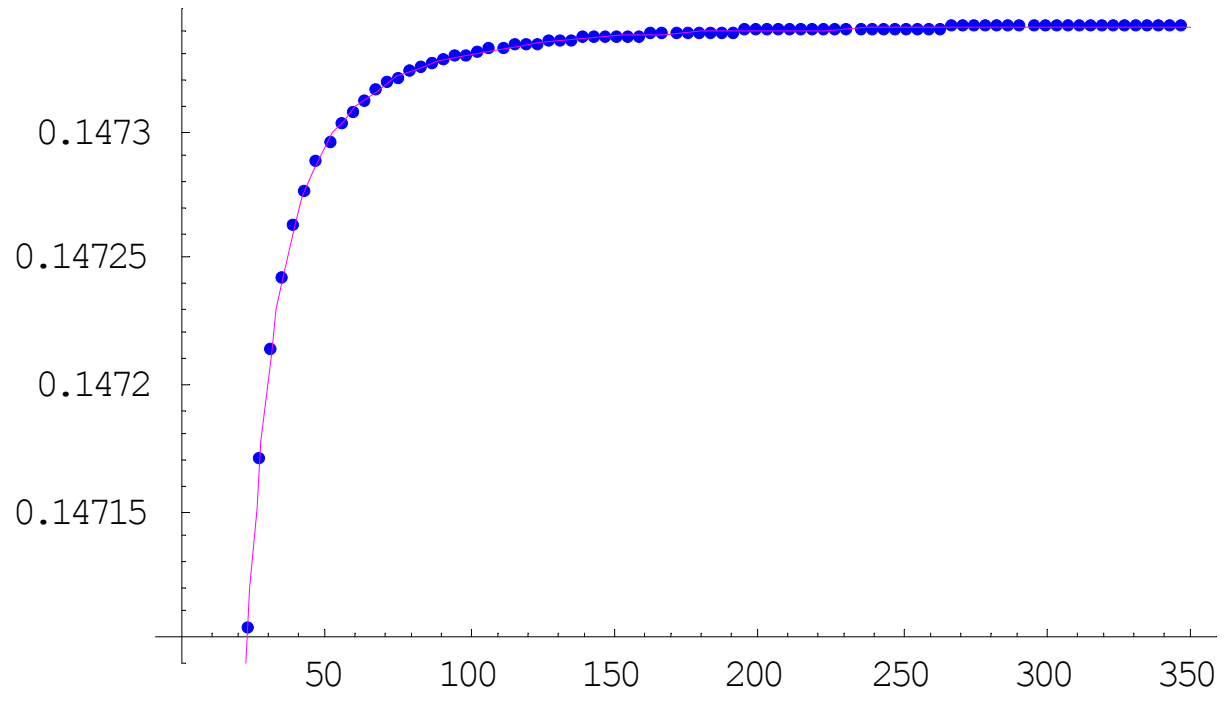

Fig.4 Comparison of the curve of nonlinear fit function and temperature value at the central point under a series of grid resolution; the blue points are data points and the thin pink line is the linear fit curve

In practical numerical simulation, there are two kinds of problems: 1. prediction of unknown questions 1. verification of known results. When predict unknown question, it should reach stage

B——grid-independent stage. When verify already known result, grid independent value $a$ should equal to the known result, or else either the numerical simulation or known result is wrong.

## 3.2 Simulation of heat transfer without heat source

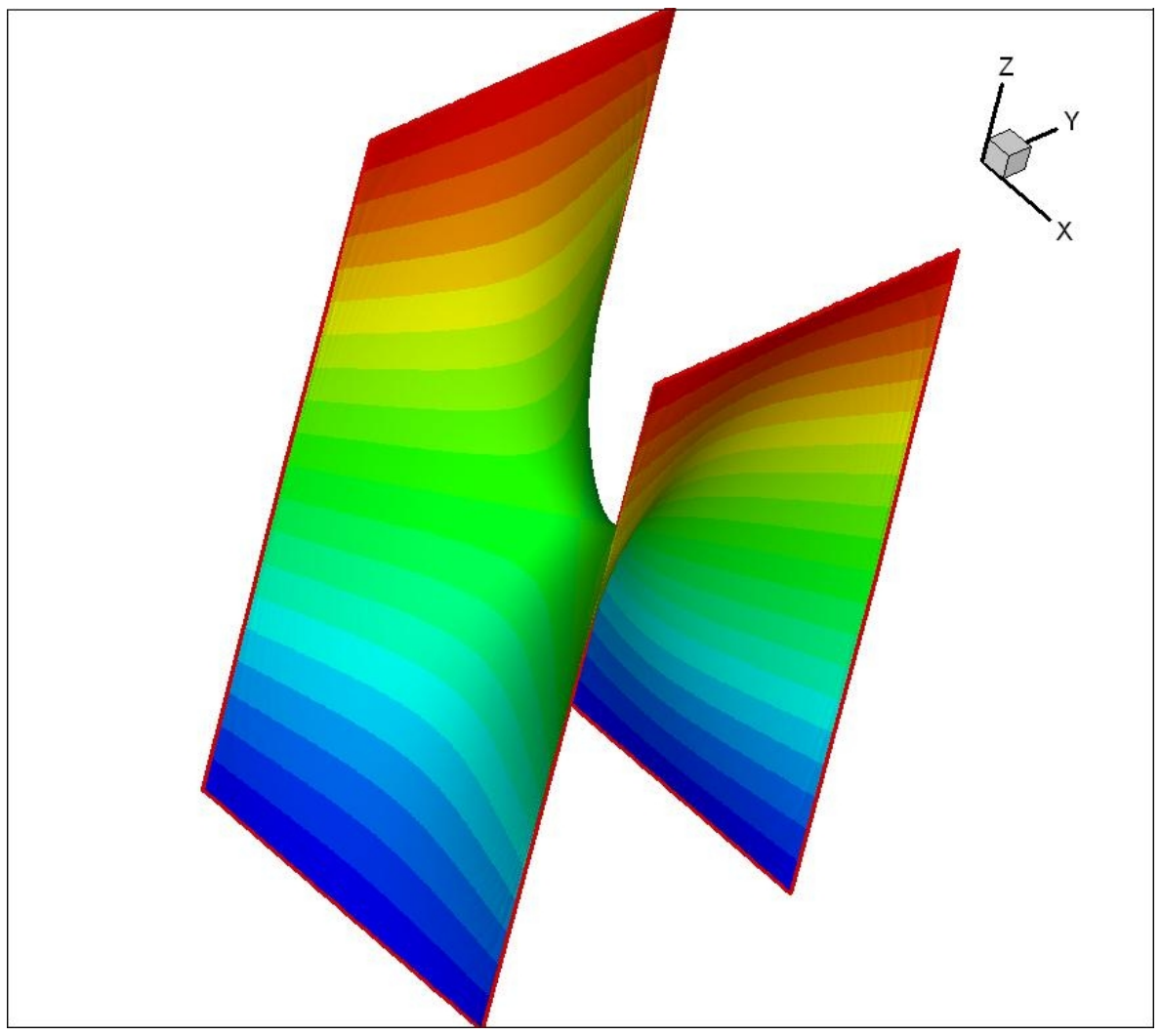

Fig. 5 Temperature field of heat transfer without heat source; Boundaries are set to different fixed temperature

Left and right boundaries are set to fixed temperature 100, front and back boundaries are set to fixed temperature 0. There is no heat source on the field. Fig 5 shows the temperature distribution on field without heat source. As shown in Fig.5 the temperature field is distorted by different boundary temperature and takes on a shape of saddle with curvature on the whole field.

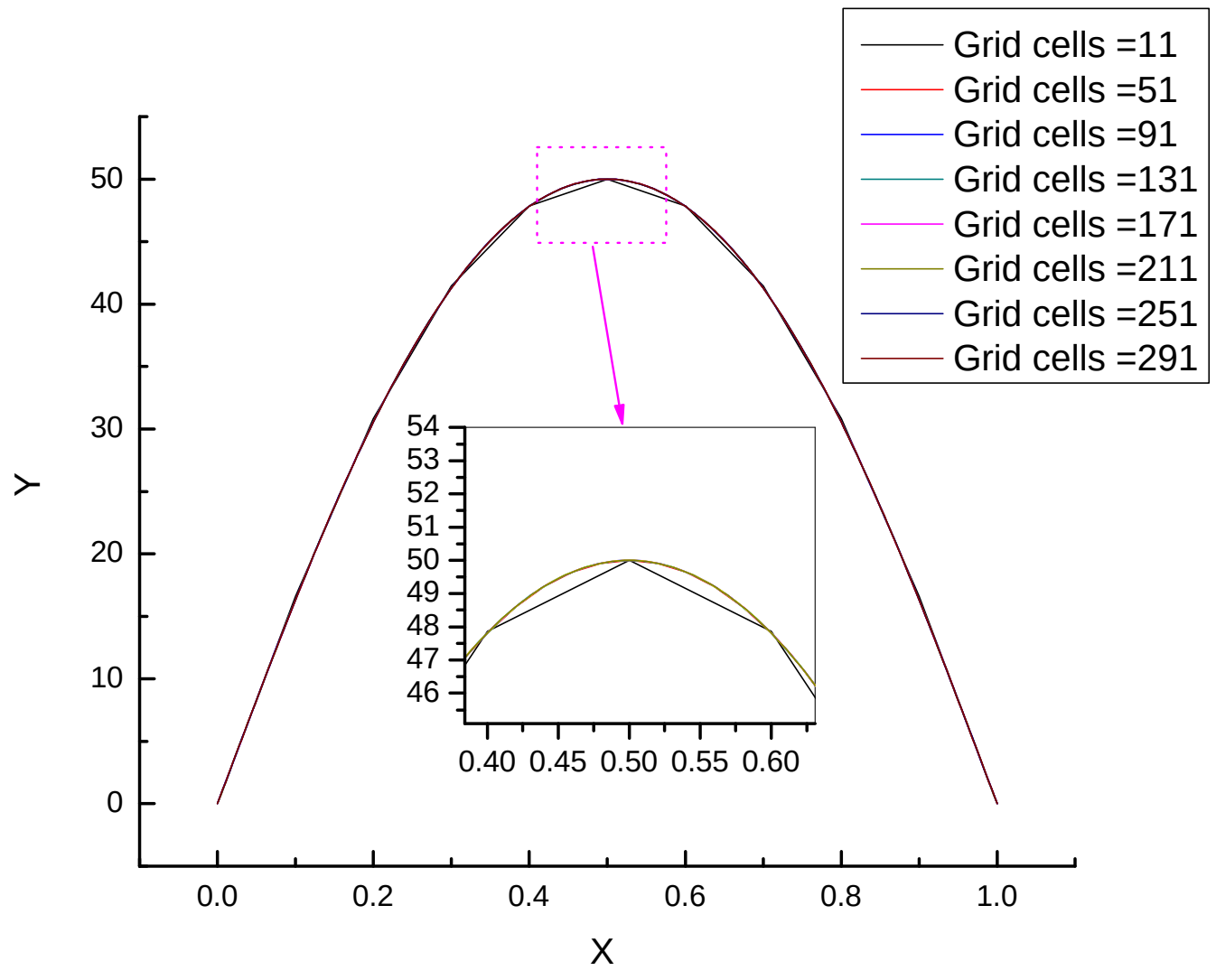

Fig. 6 Plot of temperature at the middle slice of heat transfer field under a series of number of grid cells

Fig. 6 is the middle slice of temperature field. As in Fig. 6, at each net node, the temperature equal to the exact value. At central net node, whether the grid is coarse or fine, the temperature equal to the exact value 50. It seems that grid resolution does not affect the rationality of results, but for finer grid more subtle temperature field can be achieved.

## 3.3 Simulation of heat transfer with linear distributing temperature

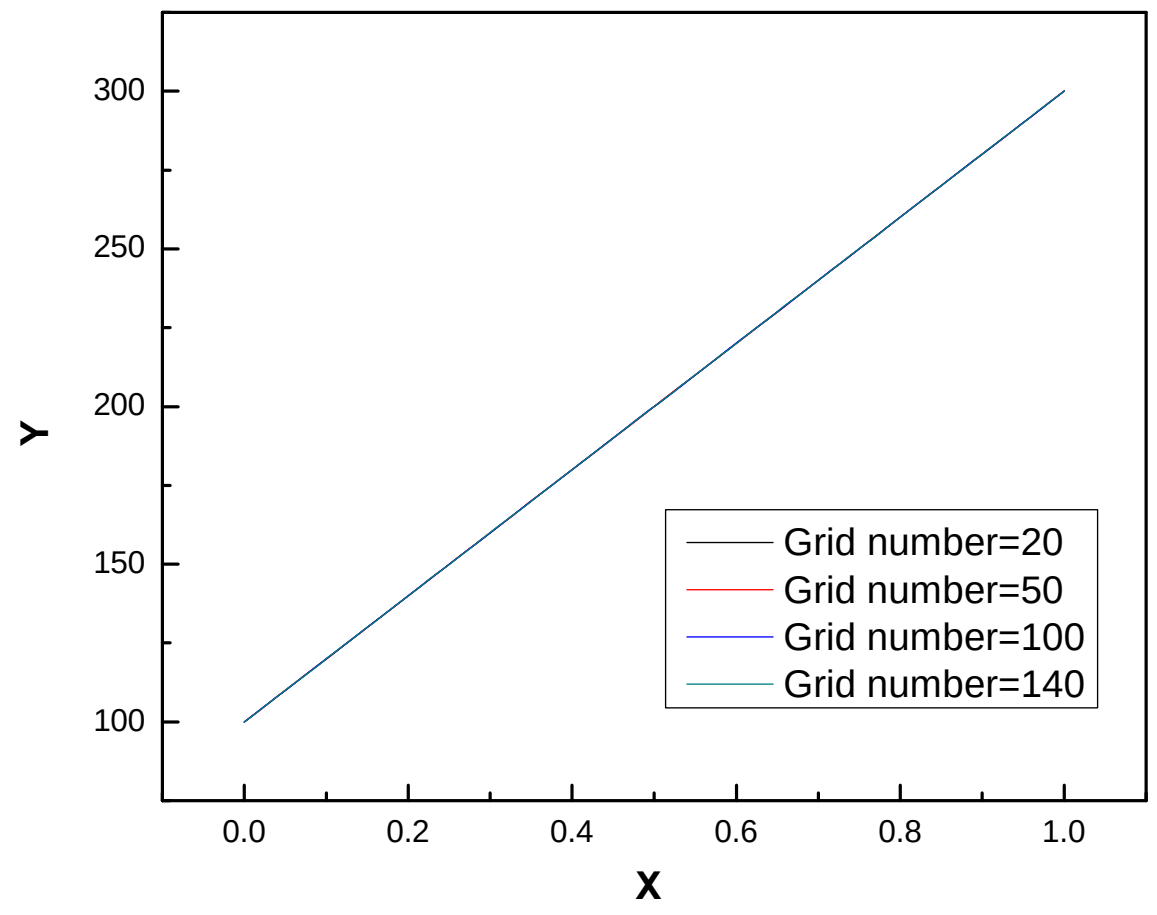

Fig. 7 Plot of streamwise temperature on the heat transfer field with linear distributing temperature under a series of number of grid cells; Front boundary is set to adiabatic; back boundary is set to fixed temperature

Left, right and front boundaries are set to adiabatic, i.e. $\frac{\partial T}{\partial n} = 0$. Back boundary is set to fixed temperature 0. Since left and right boundaries are set to adiabatic, the spanwise (from left to right) temperature distribution is uniform. Fig. 7 is the streamwise (from front to back) temperature distribution, which shows a linear distribution. The result when grid is 20×20 is as same as the result when grid is 140×140 or finer. Even coarsest grid, we can still get the exact temperature on the whole field.

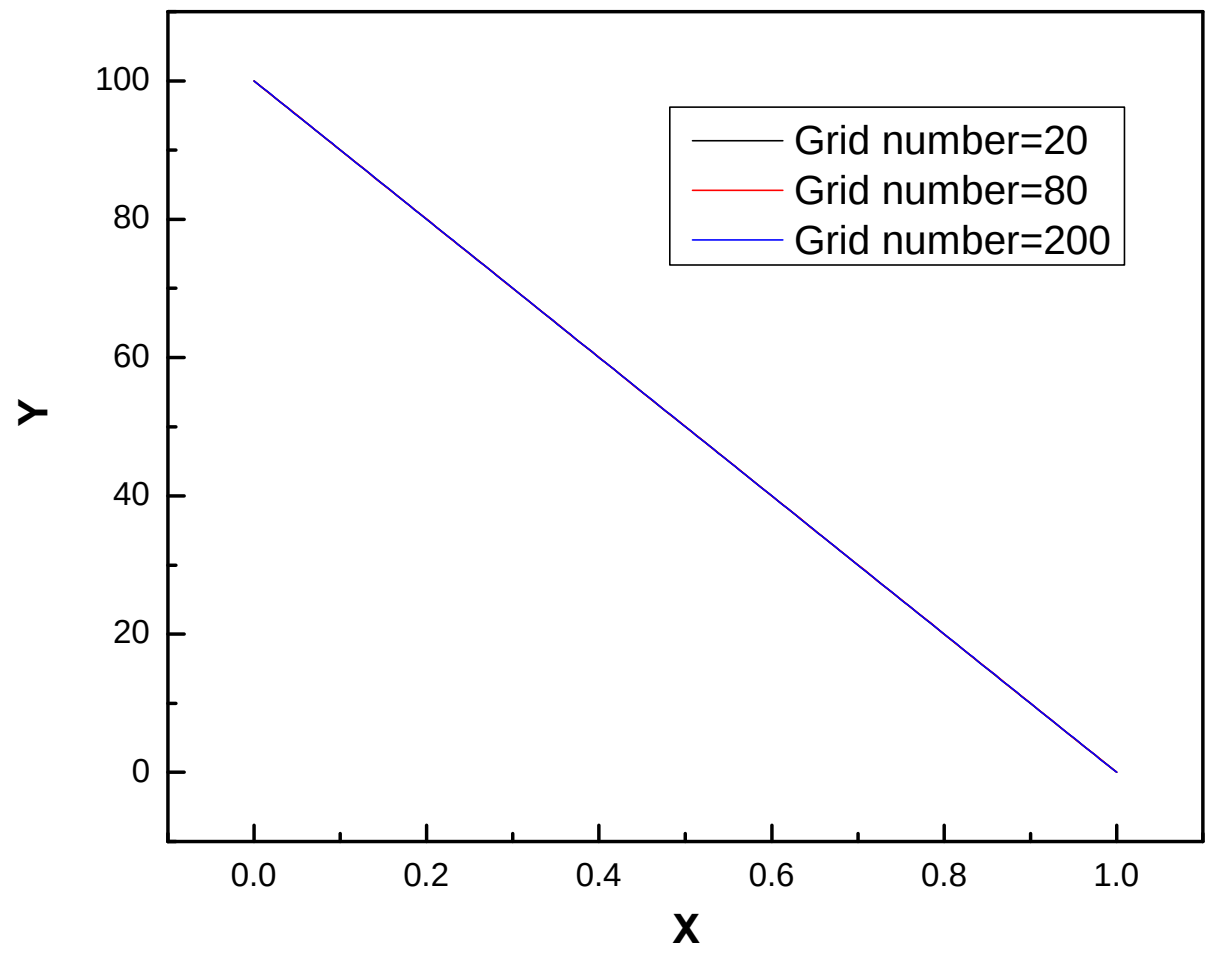

Fig. 8 Plot of streamwise temperature on the heat transfer field with linear distributing temperature under a series of number of grid cells; Front and boundaries are set to fixed temperature

Left and right boundaries are set to adiabatic. Front boundary is set to fixed temperature 100 and back boundary is set to fixed temperature 0. The spanwise (from left to right) temperature is uniform. Fig. 8 presents the streamwise (from front to back) distribution of temperature. From Fig. 8 we can see that the temperature decreases from 100 at front boundary to 0 at back boundary with a linear variation. The streamwise distribution of temperature takes on linear relation. Whether the grid is as coarse as 20×20 or fine as 200×200， the result is the same. On the whole field we get the exact temperature with arbitray number of grid cells.

As to heat transfer with linear temperature distribution, grid resolution has no effect on the result of numerical simulation.

# 4 Conclusion

This study reveals in what ways the numerical results depend on grid resolution in numerical heat transfer.

1) For numerical simulation of heat transfer with heat source, the variation of temperature against grid resolution accords with a particular curve. Two stages can be divided in the variation. Numerical results in the second stage (stage B) are considered to be grid-independent results. The law of grid dependence is given as (10). By this mathematical formula, grid independent value can be obtained by part coarse grid simulation results. Tremendous computational resource will be saved and this is of great benefit to numerical simulation.
2) For numerical simulation of heat transfer without heat source, grid resolution does not affect the rationality of results, but for finer grid more subtle temperature field can be achieved.
3) For numerical simulation of heat transfer with linear distributing temperature, grid resolution has no effect on the result of numerical simulation.

4) Theoretically, grid-independent results should equal to the exact solution of governing equations in a numerically-clean environment. This can be used as a criterion to validate numerical methods and mathematical model.